\documentclass[aps,prd,showpacs,onecolumn,preprintnumbers,notitlepage,
nofootinbib,amsfont,amssymb,10pt,singlespacing]{revtex4-1}
\usepackage{amsmath}
\usepackage{bm}
\usepackage{times}
\usepackage{braket}
\usepackage{color}
\usepackage{epsfig}
\usepackage{slashed}
\usepackage{hyperref}

\newcommand{\beq}{\begin{eqnarray}}
\newcommand{\eeq}{\end{eqnarray}}

\def\lsim{ {\ \lower-1.2pt\vbox{\hbox{\rlap{$<$}\lower6pt\vbox{\hbox{$\sim$}
}}}\ } }
\def\gsim{ {\ \lower-1.2pt\vbox{\hbox{\rlap{$>$}\lower6pt\vbox{\hbox{$\sim$}
}}}\ } }
\def \epjc{ Eur. Phys. J. C }

\def \jpg{  J. Phys. G }

\def \plb{  Phys. Lett. B }

\def \prd{  Phys. Rev. D }
\def \prl{  Phys. Rev. Lett.  }
\def \zpc{  Z. Phys. C  }
\def \jhep{ J. High Energy Phys.  }

\def \ijmpe { Int. J. Mod. Phys. E }
\def \mpla{ Mod. Phys. Lett. A }

\definecolor{Red}{rgb}{1.,0.,0.}

\definecolor{Blue}{rgb}{0.,0.,1.}

\definecolor{nicered}{rgb}{0.7,0.1,0.1}
\definecolor{nicegreen}{rgb}{0.1,0.5,0.1}

\hypersetup{colorlinks,citecolor=nicegreen,linkcolor=nicered}

\begin{document}

\title{Transverse-momentum-dependent wave functions with Glauber gluons
in $B\to\pi\pi$, $\rho\rho$ decays}

\author{Xin~Liu}
\email[Electronic address: ]{liuxin@jsnu.edu.cn}
\affiliation{School of Physics and Electronic Engineering,
 Jiangsu Normal University, Xuzhou, Jiangsu 221116, People's Republic of China}

\author{Hsiang-nan~Li}
\email[Electronic address: ]{hnli@phys.sinica.edu.tw}
\affiliation{ Institute of Physics, Academia Sinica, Taipei, Taiwan 115, Republic of China;
\\
Department of Physics, Tsing-Hua University, Hsinchu, Taiwan 300, Republic of China;
\\
and Department of Physics, National Cheng-Kung University, Tainan, Taiwan 701, Republic of China}

\author{Zhen-Jun~Xiao}
\email[Electronic address: ]{xiaozhenjun@njnu.edu.cn}
\affiliation{Department of Physics and Institute of Theoretical
Physics, Nanjing Normal University, Nanjing, Jiangsu 210023,
People's Republic of China}

\date{\today}

\begin{abstract}

We investigate the Glauber-gluon effect on the $B\to\pi\pi$ and $\rho\rho$ decays,
which is introduced via a convolution of a universal Glauber phase factor
with transverse-momentum-dependent (TMD) meson wave functions in the $k_T$
factorization theorem. For an appropriate parametrization of the Glauber phase,
it is observed that a TMD wave function for the pion ($\rho$ meson) with a
weak (strong) falloff in parton transverse momentum $k_T$ leads to
significant (moderate) modification of the $B^0\to\pi^0\pi^0$ ($B^0\to\rho^0\rho^0$)
branching ratio: the former (latter) is enhanced (reduced) by about a factor of 2 (15\%).
This observation is consistent with the dual role of the pion as a massless Nambu-Goldstone
boson and as a $q\bar q$ bound state, which requires a tighter spatial distribution
of its leading Fock state relative to higher Fock states. The agreement between the
theoretical predictions and the data for all the $B \to\pi\pi$ and $\rho^0\rho^0$
branching ratios is then improved simultaneously, and it is possible to resolve
the $B\to\pi\pi$ puzzle.

\end{abstract}

\pacs{13.25.Hw, 12.38.Bx, 11.10.Hi}

\maketitle

\section{INTRODUCTION}

The large observed $B^0\to\pi^0\pi^0$ branching ratio has been known as
a puzzle in two-body hadronic $B$ meson decays, whose data
\footnote{The latest measurement of ${\cal B}(B^0 \to \pi^0 \pi^0)=
(0.90 \pm 0.12 \pm 0.10)\times 10^{-6}$ with 6.7$\sigma$ was released
by the Belle Collaboration at ICHEP2014 \cite{Petric:2014}.} \cite{HFAG},
\begin{eqnarray}
{\cal B}(B^0 \to \pi^0 \pi^0) &=& \left\{ \begin{array}{lll}
(1.83 \pm 0.21 \pm 0.13)\times 10^{-6}\;\;\;\;({\rm BABAR}),&  \\
(0.90 \pm 0.12 \pm 0.10)\times 10^{-6}\;\;\;\;\;({\rm Belle}) ,&  \\
(1.17 \pm 0.13)\times 10^{-6}\;\;\;\;\hspace{1.08cm}({\rm HFAG}), &   \\ \end{array} \right.
\label{data}
\end{eqnarray}
show discrepancy with the predictions obtained in the perturbative QCD
(PQCD) \cite{LUY} and QCD-improved factorization (QCDF)
\cite{BY05} approaches. In resolving this puzzle, one must consider the
constraint from the $B^0\to\rho^0\rho^0$ data,
\begin{eqnarray}
{\cal B}(B^0 \to \rho^0 \rho^0) &=& \left\{ \begin{array}{lll}
(0.92 \pm 0.32 \pm 0.14)\times 10^{-6}\;\;\;\;({\rm BABAR})\;,&  \\
(1.02 \pm 0.30 \pm 0.15)\times 10^{-6}\;\;\;\;\;({\rm Belle})\;,&  \\
(0.97 \pm 0.24)\times 10^{-6}\;\;\;\;\hspace{1.08cm}({\rm HFAG})\;, &   \\ \end{array} \right.
\label{data2}
\end{eqnarray}
which, similar to the $B^0\to\pi^0\pi^0$ ones, are
dominated by the color-suppressed tree amplitude $C$. We have carefully
investigated the $B\to\pi\pi$ puzzle in the PQCD approach
based on the $k_T$ factorization theorem \cite{LUY,KLS} by calculating the
subleading contributions to the amplitude $C$. It was found
that the next-to-leading-order (NLO) contributions from the vertex
corrections, the quark loops and the magnetic penguin increased
$C$, and accordingly, they increased the $B^0\to\pi^0\pi^0$ branching ratio from
the leading-order (LO) value $0.12\times 10^{-6}$ to $0.29\times 10^{-6}$
\cite{LMS05}. At the same time, these NLO corrections increased
the $B^0\to\rho^0\rho^0$ branching ratio from the LO value
$0.33\times 10^{-6}$ to $0.92\times 10^{-6}$ \cite{LM06}, which is
 consistent with the data in Eq.~(\ref{data2}). Although the latest updates
\cite{Zhang:2014bsa} of the $B \to \pi \pi$ analysis in the PQCD formalism have
included all currently known NLO contributions, in particular, those
to the $B \to \pi$ transition form factors \cite{Li:2012nk}, the agreement between the
theoretical predictions and the data is still not satisfactory. That is, the
$B^0\to\rho^0\rho^0$ data can be easily understood in PQCD \cite{LM06} and
QCDF \cite{BRY06}, but the $B^0\to\pi^0\pi^0$ data cannot.

The different phenomenological implication of the $B\to\pi\pi$ and $\rho\rho$ data
has been noticed in the viewpoint of isospin triangles
\cite{BLS0602}, which stimulated the proposal of a new isospin
amplitude with $I = 5/2$ for the latter. It has been argued \cite{KV07} that the final-state
interaction (FSI) \cite{CHY,Chua08} could enhance the $B^0\to\pi^0\pi^0$
branching ratio through the $\rho\rho\to\pi\pi$ chain.
The $B^0\to\rho^0\rho^0$ branching ratio was not affected, since
the $\pi\pi\to\rho\rho$ chain is less important due to the smaller
$B\to\pi\pi$ branching ratios. However,
the $\rho\rho\to\rho\rho$ chain via the $t$-channel $\rho$-meson
exchange was not taken into account in the above analysis. In fact,
the $\rho$-$\rho$-$\rho$ coupling is identical to the
$\rho$-$\pi$-$\pi$ coupling in the chiral limit \cite{DSGJ}, whose
inclusion will increase the $B^0\to\rho^0\rho^0$ branching ratio,
and overshoot the data. Besides, the $\rho\rho\to\pi\pi$ chain
is expected to enhance the $B^0\to\pi^+\pi^-$ branching ratio, which
already saturates the data in the factorization theorems \cite{LMS05,BY05}.
Possible new physics signals from the $B\to\pi\pi$ decays have been discussed in
\cite{BBLS,YWL2,CGHW}. Similarly, a new-physics mechanism employed
to resolve the $B\to\pi\pi$ puzzle usually contributes to the $B\to\rho\rho$
decays, and is strongly constrained. It has been elaborated \cite{LM06} that there
is no satisfactory resolution in the literature: the subleading corrections in the
factorization theorems \cite{LMS05,LM06,BY05,BRY06,CL12} do not survive
the constraints from the $B\to\rho\rho$ data, and other resolutions are data fitting,
such as those by means of the jet function in the soft-collinear effective theory
\cite{BPS05} and the model-dependent FSI \cite{CHY,Chua08,CCS,KV07}.

It is crucial to explore any mechanism that could lead to
different color-suppressed tree amplitudes in the $B^0\to\pi^0\pi^0$
and $\rho^0\rho^0$ decays, and to examine whether it can resolve the
$B\to\pi\pi$ puzzle. We have identified a new type of infrared divergence
called the Glauber gluons \cite{CQ06}, from higher-order corrections
to the spectator diagrams in two-body hadronic $B$ meson decays \cite{LM11}.
These residual divergences were observed in the $k_T$
factorization theorem for complicated inclusive processes,
such as hadron hadroproduction \cite{CQ06}. They also appear
in the $k_T$ factorization for $B\to M_1M_2$ decays, with
the $M_2$ meson being emitted from the weak vertex, which
are dominated by contributions from the end-point region
of meson momentum fractions. The all-order summation of
the Glauber gluons, coupling the $M_2$ meson and the
$B\to M_1$ transition form factor, generates a phase factor
written as the expectation value of two transversely
separated lightlike path-ordered Wilson lines \cite{CL09}.  It is
noticed that the Glauber factor constructed in \cite{CL09} is similar to the
transverse-momentum broadening factor for an energetic parton propagating
through quark-gluon plasma \cite{LRW07,BBEV12}. The phase factor associated with
$M_2$ modifies the interference between the spectator diagrams for $C$. We
postulated that only the Glauber effect from a pion is significant, due to its
special role as a pseudo Nambu-Goldstone (NG) boson and as a $q\bar q$ bound state
simultaneously \cite{NS08}. It was then demonstrated that by tuning
the Glauber phase, the magnitude of $C$ was increased, and the
$B^0\to\pi^0\pi^0$ branching ratio could reach $1.0\times 10^{-6}$
\cite{LM11}. A thorough analysis of $B\to M_1M_2$ decays has been
carried out recently, and the Glauber gluons coupling the $M_1$ meson and
the $B\to M_2$ system were also found \cite{Li:2014haa}. The resultant phase
factor modifies the interference between the enhanced $C$ and the
color-allowed tree amplitude $T$. It turns out that the NLO PQCD prediction
for the $B^+\to\pi^+\pi^0$ branching ratio, which receives contributions from
both $T$ and $C$, also becomes closer to the data.

However, the Glauber phase in \cite{LM11,Li:2014haa} was treated as a free
parameter, so it is not clear how important this phase could be. The
postulation on the uniqueness of the pion relative to other mesons
is also lacking quantitative support. According to \cite{LM11}, the
Glauber phase factor is universal, depends on the transverse momenta
$l_T$ of Glauber gluons, and appears in a convolution with decay amplitudes
in the $k_T$ factorization theorem. Therefore, the universal phase factor
produces different Glauber effects through convolutions with the
distinct transverse-momentum-dependent (TMD) meson wave functions.
To verify this conjecture, we parametrize the universal phase factor
associated with the $M_1$ and $M_2$ mesons as a function of the variable $b$
conjugate to $l_T$, which denotes the transverse separation between
the two lightlike Wilson lines mentioned above \cite{CL09}. The convolutions
of this phase factor with the TMD pion and $\rho$ meson wave functions
proposed in \cite{Huang80}, which exhibit a weaker falloff and a stronger
falloff in the parton transverse momentum $k_T$, respectively, indicate that
the Glauber effect is indeed more significant in the $B\to\pi\pi$ decays than
in the $B\to\rho\rho$ decays. This observation is consistent with the dual
role of the pion as a massless NG boson and as a $q\bar q$ bound
state, which requires a tighter spatial distribution of its leading Fock
state relative to higher Fock states \cite{NS08}. The predicted $B^0\to\pi^0\pi^0$ and
$B^+\to\pi^+\pi^0$ branching ratios in NLO PQCD then reach $0.61 \times 10^{-6}$
from $0.29 \times 10^{-6}$ and $4.45\times 10^{-6}$ from $3.35\times 10^{-6}$,
respectively. The $B^0\to\pi^+\pi^-$ branching ratio
decreases from $6.19\times 10^{-6}$ to $5.39\times 10^{-6}$. Employing the same framework,
we obtain the $B^0\to\rho^0\rho^0$ branching ratio slightly reduced from $1.06\times 10^{-6}$
to $0.89\times 10^{-6}$. It is obvious that the agreement between the NLO PQCD predictions
and the data is greatly improved for all the above modes.

We establish the $k_T$ factorization of the $B\to\pi\pi$ and $\rho\rho$
decays including the Glauber phase factors associated with the $M_1$ and
$M_2$ mesons in Sec.~II. Section~III contains the parametrizations of the
universal Glauber phase factor, and of the intrinsic $k_T$ dependencies of the
pion and $\rho$ meson wave functions. Numerical results together with
theoretical uncertainties in our calculations are presented.
Section~IV is the conclusion.

\section{FACTORIZATION FORMULAS}

\begin{figure}[t]
\begin{center}
\includegraphics[height=3.5cm]{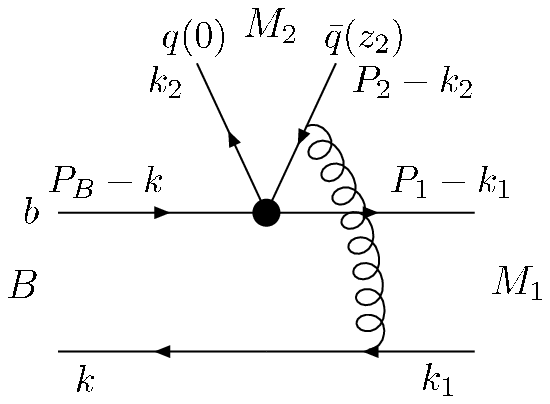}\hspace{1.0cm}
\includegraphics[height=3.5cm]{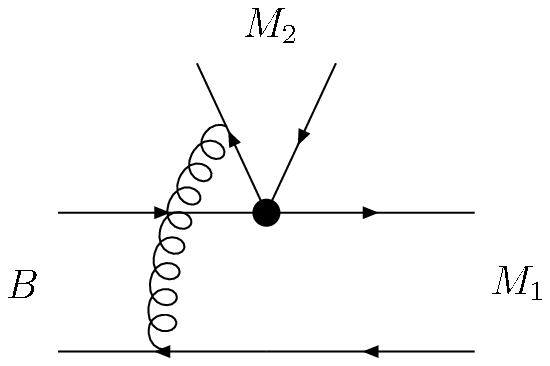}

(a)\hspace{5.5cm}(b)
\caption{LO spectator diagrams for the $B \to M_1 M_2$ decay.} \label{fig1}
\end{center}
\end{figure}

In this section we derive the PQCD factorization formulas for the
$B(P_B)\to M_1(P_1)M_2(P_2)$ decay, in which the Glauber-gluon
effect is taken into account. The $B$ meson, $M_1$ meson, and $M_2$ meson momenta
are labeled by $P_B$, $P_1$, and $P_2$, respectively, for which we
choose $P_B=(P_B^+,P_B^-,{\bf 0}_T)$ with $P_B^+=P_B^-=m_B/\sqrt{2}$, $m_B$
being the $B$ meson mass, and $P_1$ ($P_2$) in the plus (minus) direction.
The parton four-momenta $k$, $k_1$, and $k_2$ are carried by the spectator
of the $B$ meson, by the spectator of the $M_1$ meson, and by the valence
quark of the $M_2$ meson, respectively, as labeled in Fig.~\ref{fig1}.
Specifically, we keep $k^-=xP_B^-$, $k_1^+=x_1P_1^+$, $k_2^-=x_2P_2^-$, and
transverse components in hard kernels for $b$-quark decays. For the detailed
analysis of the Glauber divergences associated
with the $M_1$ and $M_2$ mesons, refer to Ref.~\cite{Li:2014haa}.

\subsection{Glauber gluons from $M_2$ meson}

\begin{figure}[b]
\begin{center}
\includegraphics[height=3.0cm]{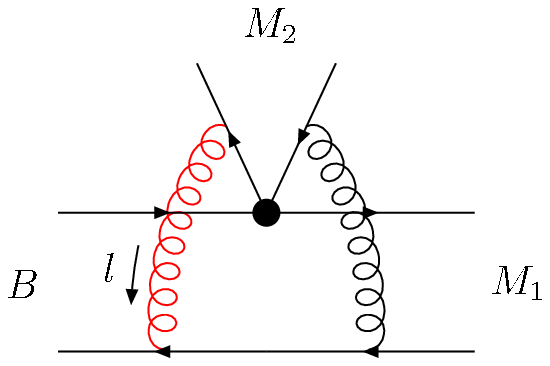}\hspace{1.0cm}
\includegraphics[height=3.0cm]{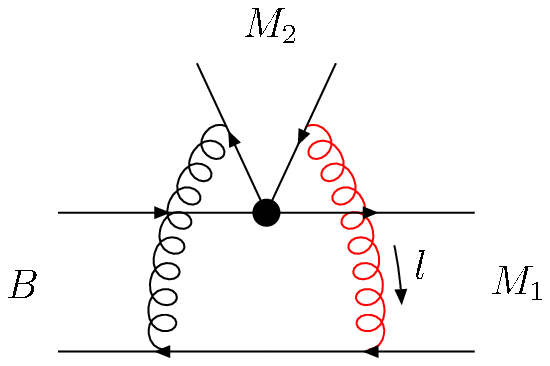}

(a)\hspace{5.0cm}(b)
\caption{(Color online) NLO spectator diagrams for the $B \to M_1 M_2$ decay
that contain the Glauber divergences associated with the $M_2$ meson.
Other NLO diagrams with the Glauber divergences can be found in
\cite{LM11}.} \label{fig2}
\end{center}
\end{figure}

\begin{figure}[t]
\begin{center}
\begin{tabular}{ccc}
\includegraphics[height=4.5cm]{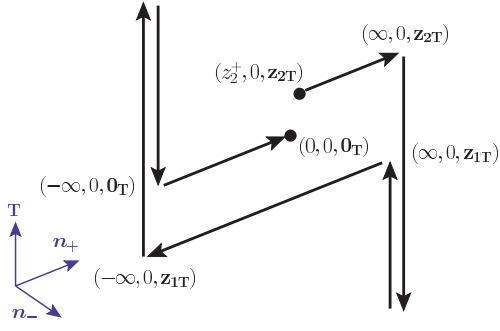}
\end{tabular}
\caption{Wilson links for the modified wave function $\phi_{2}^G$.}\label{wilson}
\end{center}
\end{figure}

We formulate the amplitude from Fig.~\ref{fig1}(a) for
the $B\to M_1 M_2$ decay in the
presence of the Glauber divergences, in which the hard gluon is
exchanged on the right, and the Glauber gluon is exchanged on the
left as shown in Fig.~\ref{fig2}(a). The spectator propagator on the $B$ meson
side can be approximated by the eikonal propagator proportional to
$-1/(l^-+i\epsilon)$ as $l$ is collinear to $P_2$, which
contains an imaginary piece $i\pi\delta(l^-)$. The propagators
of the valence antiquark and quark, with the momenta $P_2-k_2-k+k_1-l$
and $k_2+l$, respectively, generate poles on the opposite half-planes
of $l^+$ as $l^-=0$. That is, the contour integration over $l^+$
does not vanish, and the Glauber gluon with the invariant mass
$-l_T^2$ contributes a logarithmic infrared divergence $\int d^2l_T/l_T^2$
around $l_T\to 0$. Since a Glauber gluon is spacelike, and we are analyzing exclusive
processes, no real gluon emissions, such as the rung gluons
in the Balitsky-Fadin-Kuraev-Lipatov ladder \cite{BFKL},
are considered. Including the additional Glauber divergences, we propose the Wilson
links described in Fig.~\ref{wilson} for the modified $M_2$ meson
wave function, which are motivated by the observation made in \cite{LW14}:
it runs from $z_2$ to plus infinity along the $n_+$
direction, along the transverse direction to infinity and then back
to ${\bf z}_{1T}$ (the transverse coordinate of the spectator quark
in the $M_1$ meson), from plus infinity to minus infinity along
$n_+$ at the transverse coordinate ${\bf z}_{1T}$, along the
transverse direction to infinity and then back to the zero
transverse coordinate, and at last back to the origin from minus
infinity at the zero transverse coordinate. Moving the Wilson link,
which runs from plus infinity to minus infinity along $n_+$, to
${\bf z}_{1T}\to \infty$, we obtain the standard $M_2$ meson wave
function \cite{LW14} without the Glauber divergences. This
Wilson link at the finite transverse coordinate
${\bf z}_{1T}$ leads to the $\delta(l^-)$ function.

The modified $M_2$ meson wave function depends on two
transverse coordinates ${\bf z}_{1T}$ and ${\bf z}_{2T}$, denoted as
$\phi_{2}^G({\bf z}_{1T},{\bf z}_{2T})$, where the dependence on
$z_2^+$ has been suppressed. It has been shown that the Glauber gluon in the
$B\to M_1M_2$ decay can be further factorized from the $M_2$ meson
in the dominant kinematic region, and summed to all orders into
a phase factor $G({\bf z}_{1T}-{\bf z}_{T})$. We then have the convolution
\begin{eqnarray}
\phi_{2}^G({\bf z}_{1T},{\bf z}_{2T})=\int d^2{\bf z}_T
G({\bf z}_{1T}-{\bf z}_{T}){\bar\phi}_{2}({\bf z}_{T},{\bf z}_{2T}+{\bf z}_{T}),
\label{con}
\end{eqnarray}
where the definition for the two-coordinate wave function
${\bar\phi}_{2}({\bf z}_{T},{\bf z}_{2T})$, similar to that in
\cite{CL09}, will be given in Eq.~(\ref{rm2}) below.
The Wilson lines of $G({\bf z}_{1T}-{\bf z}_{T})$
contain the longitudinal piece, which runs from minus infinity to
plus infinity in the direction $n_-=(0,1,{\bf 0}_T)$ at the
transverse coordinate ${\bf z}_{T}$ \cite{CL09}, in addition to the
longitudinal piece at the transverse coordinate ${\bf z}_{1T}$ in
Fig.~\ref{wilson}. The above Wilson links are similar to that constructed
for the jet quenching parameter in \cite{BBEV12}, which is
defined as the average transverse momentum squared with respect to
the original direction of motion that a highly energetic parton
picks up, while traveling through a nuclear medium.
If the Glauber factor contributes only a constant
phase, Eq.~(\ref{con}) reduces to \cite{CL09}
\begin{eqnarray}
\phi_{2}^G({\bf z}_{1T},{\bf z}_{2T})&\approx&\exp(iS_{e2})\int
d^2{\bf z}_T {\bar\phi}_{2}({\bf z}_{T},{\bf z}_{2T}+{\bf z}_{T}),\nonumber\\
&\equiv&\exp(iS_{e2})\phi_{2}({\bf z}_{2T}),\label{glau}
\end{eqnarray}
where $\phi_{2}({\bf z}_{2T})$ denotes the standard $M_2$ meson wave function.
The approximation in Eq.~(\ref{glau}) with the constant Glauber phase has been
adopted in \cite{LM11,Li:2014haa}.

We route the transverse loop momentum ${\bf l}_T$ of the Glauber gluon
through the hard gluon, the valence antiquark of the $M_2$ meson, and
the valence quark of the $M_2$ meson in Fig.~\ref{fig1}(a). Regarding the
Glauber gluon, the valence quark, and the valence antiquark as the partons
of the $M_2$ meson, we assign $-{\bf l}_T$, ${\bf k}_{2T}$, and $-{\bf k}_{2T}+{\bf l}_T$
to them, respectively. That is, the set of the Glauber gluon, the valence quark,
and the valence antiquark does not carry net transverse momenta.
The corresponding amplitude is modified into
\begin{eqnarray}
\int \frac{d^2{\bf k}_T}{(2\pi)^2} \frac{d^2{\bf k}_{1T}}{(2\pi)^2}
\frac{d^2{\bf k}_{2T}}{(2\pi)^2}\int\frac{d^2 {\bf l}_T}{(2\pi)^2}
\phi_B({\bf k}_T)\phi_1({\bf k}_{1T})\bar\phi_2({\bf k}_{2T},-{\bf k}_{2T}+{\bf l}_T)
G_2({\bf l}_T)H_a({\bf k}_T,{\bf k}_{1T},{\bf k}_{2T}, {\bf l}_T),\label{1}
\end{eqnarray}
where the convolution in momentum fractions has been suppressed, and
$\phi_B$, $\phi_1$, and $H_a$ denote
the $B$ meson wave function, the $M_1$ meson wave function,
and the hard $b$-quark decay kernel, respectively. The Glauber factor $G_2({\bf l}_T)$
in momentum space appears as an additional convolution piece in the PQCD factorization
formula for Fig.~\ref{fig1}(a).

The virtual gluon and the virtual quark in the hard kernel $H_a$ have
the transverse momenta ${\bf k}_T+{\bf l}_T-{\bf k}_{1T}$ and ${\bf k}_{1T}-{\bf k}_{2T}-{\bf k}_T$,
respectively. We apply the variable changes ${\bf k}_{1T}-{\bf l}_T\to
{\bf k}_{1T}$ and ${\bf k}_{2T}-{\bf l}_T\to {\bf k}_{2T}$, such that ${\bf l}_T$
flows through the spectator quark in the $M_1$ meson, the valence quark in the $M_1$
meson, and the valence quark in the $M_2$ meson. Then the ${\bf l}_T$
dependence disappears from the hard kernel, and Eq.~(\ref{1}) becomes
\begin{eqnarray}
\int \frac{d^2{\bf k}_T}{(2\pi)^2} \frac{d^2{\bf k}_{1T}}{(2\pi)^2}
\frac{d^2{\bf k}_{2T}}{(2\pi)^2}\int\frac{d^2{\bf l}_T}{(2\pi)^2}
\phi_B({\bf k}_T)\phi_1({\bf k}_{1T}+{\bf l}_T)\bar\phi_2({\bf k}_{2T}+{\bf l}_T,-{\bf k}_{2T})
G_2({\bf l}_T)H_a({\bf k}_T,{\bf k}_{1T},{\bf k}_{2T}).\label{rm1}
\end{eqnarray}
We perform the Fourier transformation of Eq.~(\ref{rm1}) by
employing
\begin{eqnarray}
\phi_1({\bf k}_{1T}+{\bf l}_T)&=&\int d^2 {\bf b}_1\exp[i({\bf k}_{1T}+{\bf l}_T)\cdot {\bf b}_1]\phi_1({\bf b}_1),\\
\bar\phi_2({\bf k}_{2T}+{\bf l}_T,-{\bf k}_{2T})&=&\int d^2 {\bf b}'_2d^2 {\bf b}_2\exp[i({\bf k}_{2T}+{\bf l}_T) \cdot {\bf b}'_2]\exp[-i{\bf k}_{2T}\cdot (-{\bf b}_2-{\bf b}_1-{\bf b}')]\bar\phi_2({\bf b}'_2,{\bf b}_2+{\bf b}_1+{\bf b}'),\label{rm2}\\
G_2({\bf l}_T)&=&\int d^2 {\bf b}'\exp(i{\bf l}_T\cdot {\bf b}')\exp[i S({\bf b}')],\label{rm3}
\end{eqnarray}
where the phase factor $\exp[i S({\bf b}')]$ is a consequence of the all-order
summation of Glauber gluons in $b'$ space \cite{CL09}.
Working out the integration over ${\bf l}_T$ and ${\bf b}'_2$, and adopting
\begin{eqnarray}
H_a({\bf b}_1,{\bf b}_2)\delta^{(2)}({\bf b} -{\bf b}_1)=\int \frac{d^2{\bf k}_T}{(2\pi)^2}
\frac{d^2{\bf k}_{1T}}{(2\pi)^2} \frac{d^2{\bf k}_{2T}}{(2\pi)^2}\exp(i{\bf k}_T\cdot
{\bf b}+i{\bf k}_{1T}\cdot {\bf b}_1+i{\bf k}_{2T}\cdot {\bf b}_2)
H_a({\bf k}_T,{\bf k}_{1T},{\bf k}_{2T}),\label{h}
\end{eqnarray}
we obtain the PQCD factorization formula
with the Glauber effect from the $M_2$ meson being included,
\begin{eqnarray}
\int d^2{\bf b}_{1} d^2{\bf b}_{2} d^2{\bf b}'
\phi_B({\bf b}_1)\phi_1({\bf b}_{1}){\bar\phi}_2({\bf b}_1+{\bf b}',{\bf b}_2+{\bf b}_1+{\bf b}')\exp[i S({\bf b}')]
H_a({\bf b}_{1},{\bf b}_{2}).\label{3}
\end{eqnarray}

For Fig.~\ref{fig1}(b) with the hard
gluon being exchanged on the left, we derive
\begin{eqnarray}
\int d^2{\bf b}_{1} d^2{\bf b}_{2} d^2{\bf b}'
\phi_B({\bf b}_1)\phi_1({\bf b}_{1}){\bar\phi}_2({\bf b}_2+{\bf b}_1+{\bf b}',{\bf b}_1+{\bf b}')\exp[-i S({\bf b}')]
H_b({\bf b}_{1},{\bf b}_{2}).\label{3m}
\end{eqnarray}
Note the negative phase in the factor $\exp[-i S({\bf b}')]$, which
is attributed to the Glauber gluons emitted by the valence antiquark
of the $M_2$ meson.
Equations~(\ref{3}) and (\ref{3m}) imply that the Glauber
effect gives a strong phase to each spectator diagram for the
color-suppressed tree amplitude. It is equivalent to route ${\bf l}_T$ through the
$B$ meson wave function in Fig.~\ref{fig1}, under which the same factorization
formulas will be attained.

\subsection{Glauber gluons from $M_1$ meson}

\begin{figure}[b]
\begin{center}
\includegraphics[height=3.0cm]{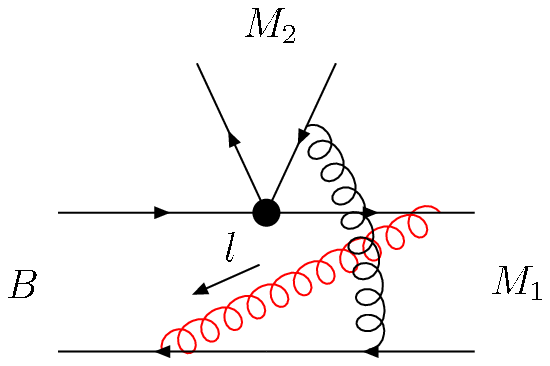}\hspace{1.0cm}
\includegraphics[height=3.0cm]{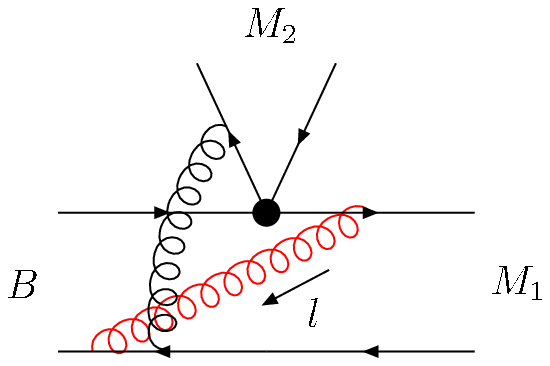}

(a)\hspace{5.0cm}(b)
\caption{(Color online) NLO spectator diagrams for the $B \to M_1 M_2$ decay
that contain the Glauber divergences associated with the $M_1$ meson.
Other NLO diagrams with the Glauber divergences are referred to
\cite{Li:2014haa}.} \label{fig4}
\end{center}
\end{figure}

We then include the Glauber gluons associated with the $M_1$
meson, starting from Fig.~\ref{fig1}(a). Some NLO diagrams that
produce these types of Glauber divergences are displayed in Fig.~\ref{fig4}.
We route the transverse loop
momentum ${\bf l}_T$ of the Glauber gluon in Fig.~\ref{fig4}(a) through the hard
gluon, the valence antiquark of the $M_2$ meson, and the valence
quark of the $M_1$ meson. The spectator propagator on the $B$ meson
side can be approximated by the eikonal propagator proportional to
$-1/(l^++i\epsilon)$ as $l$ is collinear to $P_1$, which
contains an imaginary piece $i\pi\delta(l^+)$.
The above routing of $l$ clearly indicates that the propagators
of the valence antiquark of $M_2$ and the valence quark of $M_1$,
with the momenta $P_2-k_2-k+k_1-l$
and $P_1-k_1+l$, respectively, generate poles on the opposite half-planes
of $l^-$ as $l^+=0$. That is, the contour integration over $l^-$
does not vanish, and the Glauber gluon with the invariant mass
$-l_T^2$ contributes a logarithmic infrared divergence as $l_T\to 0$.
Similarly, the all-order summation of the Glauber divergences leads to
a phase factor $G_1({\bf l}_T)$ associated with the $M_1$ meson.
The above explanation applies to the Glauber divergence in Fig.~\ref{fig4}(b),
and its all-order summation gives the same phase factor $G_1({\bf l}_T)$. The
reason is obvious from Fig.~\ref{fig4}, where the Glauber gluon always
attaches to the spectator in the $B$ meson and the valence quark in the $M_1$
meson.

Assume that the Glauber gluons from the $M_1$
meson and the $M_2$ meson carry the transverse momenta
${\bf l}_{1T}$ and ${\bf l}_{2T}$, respectively. The above
transverse momenta are routed through the mesons, instead of through the hard
kernel, so that the hard kernel has the same expression as in the
LO PQCD approach: ${\bf l}_{1T}$ flows through the
valence quark and the valence antiquark of the $M_2$ meson, and
${\bf l}_{2T}$ flows through the $M_1$ meson and then through the valence
quark of the $M_2$ meson. The resultant amplitude is written as
\begin{eqnarray}
{\cal A}_a&=&\int \frac{d^2{\bf k}_T}{(2\pi)^2} \frac{d^2{\bf k}_{1T}}{(2\pi)^2}
\frac{d^2{\bf k}_{2T}}{(2\pi)^2}\int
\frac{d^2 {\bf l}_{1T}}{(2\pi)^2}\frac{d^2{\bf l}_{2T}}{(2\pi)^2} \phi_B({\bf k}_T)
\bar\phi_1({\bf k}_{1T}+{\bf l}_{2T},-{\bf k}_{1T}-{\bf l}_{1T}-{\bf l}_{2T})\nonumber\\
& &\times\bar\phi_2({\bf k}_{2T}+{\bf l}_{1T}+{\bf l}_{2T},-{\bf k}_{2T}-{\bf l}_{1T})
G_1({\bf l}_{1T})G_2({\bf l}_{2T})H_a({\bf k}_T,{\bf k}_{1T},{\bf k}_{2T}).\label{rm5p}
\end{eqnarray}
The Fourier transformations
\begin{eqnarray}
G_1({\bf l}_{1T})&=&\int d^2 {\bf b}_{s1}\exp(i {\bf l}_{1T}\cdot
{\bf b}_{s1})\exp\left[-i S({\bf b}_{s1})\right],
\label{g1g2}
\end{eqnarray}
for the Glauber factor, and
\begin{eqnarray}
& &\phi_B({\bf k}_T)=\int d^2 {\bf b}_B\exp(i{\bf k}_T\cdot {\bf b}_B)\phi_B({\bf b}_B),\\
& &\bar\phi_1({\bf k}_{1T}+{\bf l}_{2T},-{\bf k}_{1T}-{\bf l}_{1T}-{\bf l}_{2T})\nonumber\\
&=&\int d^2 {\bf b}'_1d^2{\bf b}_1\exp[i({\bf k}_{1T}+{\bf l}_{2T})\cdot
{\bf b}'_1]\exp[-i({\bf k}_{1T}+{\bf l}_{1T}+{\bf l}_{2T})\cdot({\bf b}'_1-
{\bf b}_1)]\bar\phi_1({\bf b}'_1,{\bf b}'_1-{\bf b}_1),\\
& &\bar\phi_2({\bf k}_{2T}+{\bf l}_{1T}+{\bf l}_{2T},-{\bf k}_{2T}-{\bf l}_{1T})\nonumber\\
&=&\int d^2 {\bf b}'_2d^2{\bf b}_2\exp[i({\bf k}_{2T}+{\bf l}_{1T}+{\bf l}_{2T})\cdot
{\bf b}'_2]\exp[-i({\bf k}_{2T}+{\bf l}_{1T})\cdot({\bf b}'_2-
{\bf b}_2)]\bar\phi_2({\bf b}'_2,{\bf b}'_2-{\bf b}_2),\label{phi2ap}
\end{eqnarray}
for the meson wave functions are then inserted into Eq.~(\ref{rm5p}).
Note that the Glauber phases associated with $M_1$ and $M_2$ differ by a sign
for Fig.~\ref{fig1}(a) as shown in Eqs.~(\ref{rm3}) and (\ref{g1g2}) \cite{Li:2014haa}.

We collect the exponents depending on ${\bf l}_{1T}$ and ${\bf l}_{2T}$,
integrate them over ${\bf l}_{1T}$ and ${\bf l}_{2T}$, and obtain the
$\delta$ functions $\delta^{(2)}({\bf b}_{s1}-{\bf b}'_1+{\bf b}_1+{\bf b}_2)$ and
$\delta^{(2)}({\bf b}_{s2}+{\bf b}_1+{\bf b}'_2)$,
respectively. The next step is to perform the integration over
${\bf b}'_1$ and ${\bf b}'_2$ according to the above
$\delta$ functions, which lead to
${\bf b}'_1={\bf b}_{s1}+{\bf b}_1+{\bf b}_2$ and ${\bf b}'_2=-{\bf b}_{s2}- {\bf b}_1$.
For the ${\bf k}_T$, ${\bf k}_{1T}$, and ${\bf k}_{2T}$ integrations, we still have Eq.~(\ref{h}),
namely, ${\bf b}_B={\bf b}_1$. At last, we derive
\begin{eqnarray}
{\cal A}_a&=&\int d^2{\bf b}_{1} d^2{\bf b}_{2} \int d^2{\bf b}_{s1} d^2{\bf b}_{s2}
\phi_B({\bf b}_1)\bar\phi_1({\bf b}_{s1}+
{\bf b}_1+{\bf b}_2,{\bf b}_{s1}+{\bf b}_2)\nonumber\\
& &\times {\bar\phi}_2({\bf b}_{s2}+{\bf b}_1,{\bf b}_{s2}+{\bf b}_1+
{\bf b}_2)\exp\left[-i S({\bf b}_{s1})+i
S({\bf b}_{s2})\right] H_a({\bf b}_{1},{\bf b}_{2}),\nonumber\\
&=&\int d^2{\bf b}_{1} d^2{\bf b}_{2} \int d^2{\bf b}_{s1} d^2{\bf b}_{s2}
\bar\phi_B({\bf b}_1)\bar\phi_1({\bf b}_{s1}+
{\bf b}_1,{\bf b}_{s1})\nonumber\\
& &\times {\bar\phi}_2({\bf b}_{s2},{\bf b}_{s2}+
{\bf b}_2)\exp\left[-i S({\bf b}_{s1}-{\bf b}_2)+i
S({\bf b}_{s2}-{\bf b}_1)\right] H_a({\bf b}_{1},{\bf b}_{2}). \label{cmfp}
\end{eqnarray}
To arrive at the second expression, the variable
changes ${\bf b}_{s1}+{\bf b}_2\to {\bf b}_{s1}$ and
${\bf b}_{s2}+{\bf b}_1\to {\bf b}_{s2}$ have been employed.

For Fig.~\ref{fig1}(b) with the hard gluon exchanged on the left, we
route ${\bf l}_{1T}$ through the valence quark and the
valence antiquark of the $M_2$ meson, and route ${\bf l}_{2T}$ through the
$M_1$ meson and then through the valence antiquark of the $M_2$
meson. The resultant amplitude is factorized into
\begin{eqnarray}
{\cal A}_b&=&\int \frac{d^2{\bf k}_T}{(2\pi)^2} \frac{d^2{\bf k}_{1T}}{(2\pi)^2}
\frac{d^2{\bf k}_{2T}}{(2\pi)^2}\int
\frac{d^2{\bf l}_{1T}}{(2\pi)^2}\frac{d^2{\bf l}_{2T}}{(2\pi)^2} \phi_B({\bf k}_T)
\bar\phi_1({\bf k}_{1T}+{\bf l}_{2T},-{\bf k}_{1T}-{\bf l}_{1T}-{\bf l}_{2T})\nonumber\\
& &\times\bar\phi_2({\bf k}_{2T}-{\bf l}_{1T},-{\bf k}_{2T}+{\bf l}_{1T}+{\bf l}_{2T})
G_1({\bf l}_{1T})G_2({\bf l}_{2T})H_b({\bf k}_T,{\bf k}_{1T},{\bf k}_{2T}).\label{rm5bp}
\end{eqnarray}
The Fourier transformations are then applied with Eq.~(\ref{phi2ap}) being replaced by
\begin{eqnarray}
& &\bar\phi_2({\bf k}_{2T}-{\bf l}_{1T},-{\bf k}_{2T}+{\bf l}_{1T}+{\bf l}_{2T})\nonumber\\
&=&\int d^2 {\bf b}'_2d^2{\bf b}_2\exp[i({\bf k}_{2T}-{\bf l}_{1T})\cdot
{\bf b}'_2]\exp[-i({\bf k}_{2T}-{\bf l}_{1T}-{\bf l}_{2T})\cdot({\bf b}'_2-
{\bf b}_2)]\bar\phi_2({\bf b}'_2,{\bf b}'_2-{\bf b}_2),
\end{eqnarray}
and we have the factorization formula
\begin{eqnarray}
{\cal A}_b&=&\int d^2{\bf b}_{1} d^2{\bf b}_{2} \int d^2{\bf b}_{s1} d^2{\bf b}_{s2}
\bar\phi_B({\bf b}_1)\bar\phi_1({\bf b}_{s1}+
{\bf b}_1,{\bf b}_{s1})\nonumber\\
& &\times {\bar\phi}_2({\bf b}_{s2}+{\bf b}_2,{\bf b}_{s2})
\exp\left[-i S({\bf b}_{s1}-{\bf b}_2)-i
S({\bf b}_{s2}-{\bf b}_1)\right] H_b({\bf b}_{1},{\bf b}_{2}). \label{cmfbp}
\end{eqnarray}
The Glauber phases for the $M_1$ meson have the same sign in
${\cal A}_a$ and ${\cal A}_b$ as explained before. The expressions of the hard kernels
$H_a$ and $H_b$ from Figs.~\ref{fig1}(a) and \ref{fig1}(b), respectively, corresponding
to various tree and penguin operators, can be found in Ref.~\cite{LM11}.

\section{NUMERICAL ANALYSIS}

Recently, there were four
works~\cite{Li:2014haa,Qiao:2014lwa,Chang:2014rla,Cheng:2014rfa}
devoted to the resolution of the $B \to \pi \pi$ puzzle by enhancing the
amplitude $C$:
\begin{itemize}

\item[(a)]
In Ref.~\cite{Li:2014haa}, Li and Mishima treated the Glauber phases
as free parameters in the $B\to\pi\pi$ decays, and
postulated that they vanish in the $B\to\rho\rho$ decays. When the phases
associated with the $M_1$ and $M_2$ mesons are both chosen as $-\pi/2$ in the former,
the spectator amplitudes in the NLO PQCD formalism increase, and
the $B^0 \to\pi^0 \pi^0$ branching ratio becomes as large as
$1.2 \times 10^{-6}$.

\item[(b)]
In Ref.~\cite{Qiao:2014lwa}, Qiao {\it et al.} significantly lowered the scale for the hard spectator
interactions to the so-called optimal scale $Q_1^H \sim 0.75$ GeV in the QCDF approach
following the principle of maximum conformality, and found the $B^0 \to \pi^0 \pi^0$
branching ratio as large as $0.98^{+0.28}_{-0.32} \times 10^{-6}$.
To justify this resolution, it is crucial to examine how the $B^0 \to \rho^0 \rho^0$ branching ratio
is modified in the same analysis.

\item[(c)]
In Ref.~\cite{Chang:2014rla}, Chang {\it et al.} adopted large parameters $\rho_H$
and $\phi_H$ for the spectator amplitudes, as well as large parameters
$\rho_A$ and $\phi_A$ for the nonfactorizable annihilation ones in the QCDF framework
in order to fit the $B_{u,d} \to \pi\pi, \pi K$ and $K\bar K$ data.
As a consequence of the data fitting, they obtained extremely large $B^0 \to \pi^0 \pi^0$
branching ratios $1.67^{+0.33}_{-0.30} \times 10^{-6}$
and $2.13^{+0.43}_{-0.38}\times 10^{-6}$ corresponding to different scenarios.

\item[(d)]
In Ref.~\cite{Cheng:2014rfa}, Cheng {\it et al.} got the large color-suppressed tree amplitudes $C$ around
$0.5 e^{-i 65^\circ}$ and $0.6 e^{-i 80^\circ}$ directly through global fits to the data,
where the former arose only from the $B_{u,d} \to \pi\pi, \pi K$ and $KK$ data, while
the latter came from all the available $B_{u,d} \to PP$ data. These color-suppressed tree amplitudes
resulted in the large $B^0 \to \pi^0 \pi^0$ branching ratios $1.43 \pm 0.55 \times 10^{-6}$ and
$1.88 \pm 0.42 \times 10^{-6}$, respectively, in the framework of flavor $SU(3)$ symmetry.

\end{itemize}
The experimentally observed pattern ${\rm Br}(B^+ \to \pi^+ \pi^0) > {\rm Br}(B^0 \to \pi^+
\pi^-) > {\rm Br}(B^0 \to \pi^0 \pi^0)$ is also produced in Refs.~\cite{Li:2014haa,Qiao:2014lwa}.
The question on why the color-suppressed tree amplitudes are so different in the $B\to\pi\pi$
and $B\to\rho\rho$ decays remains to be answered.

In this section we attempt to answer this question by quantitatively estimating
the different Glauber effects in the $B\to\pi\pi$ and $\rho\rho$ decays
based on the PQCD factorization formulas in Eqs.~(\ref{cmfp}) and (\ref{cmfbp}).
As stated before, the Glauber factor is universal, namely, independent of
the final-state hadrons, because it has been factorized from the decay
processes. Nevertheless, the Glauber effect is not universal, as it
appears through the convolution with the TMD wave functions
$\bar\phi_1({\bf b}_{s1}+{\bf b}_1,{\bf b}_{s1})$ and
${\bar\phi}_2({\bf b}_{s2}+{\bf b}_2,{\bf b}_{s2})$,
which possess different intrinsic $b$ dependencies for the pion and the $\rho$
meson. It will be demonstrated that the model wave function in \cite{Huang80}
serves the purpose of revealing sufficiently distinct Glauber effects on the
$B^0\to\pi^0\pi^0$ and $B^0\to\rho^0\rho^0$ branching ratios.

\subsection{Parametrizations}

The intrinsic $k_T$ dependence of a TMD meson wave function
is usually parametrized through the factor \cite{Huang80,BHP80}
\begin{eqnarray}
{\cal M}^2=\frac{k_T^2+m^2}{x}+\frac{k_T^2+m^2}{1-x},\label{mm}
\end{eqnarray}
where $m=m_u=m_d$ denotes the constituent quark mass, and $x$ denotes the parton
momentum fraction. Below we shall drop $m^2$ for simplicity. In the collinear
factorization theorem one integrates a TMD wave function over $k_T$ to obtain
a distribution amplitude. Assume that the intrinsic $k_T$ dependence appears
in a Gaussian form \cite{YXM07},
\begin{eqnarray}
\phi_M({\bf k}_T)=\frac{\pi}{2\beta_M^2}
\exp\left(-\frac{{\cal M}^2}{8\beta_M^2}\right)\frac{\phi_M(x)}{x(1-x)},
\label{mod3}
\end{eqnarray}
where $\beta_M$ is a shape parameter for $M=\pi$ and $\rho$,
and $\phi_M(x)$ denotes the standard twist-2 and twist-3 light-cone distribution
amplitudes. Regarding the first (second) $k_T$ in Eq.~(\ref{mm}) as the
transverse momentum carried by the valence quark (antiquark) of the momentum
fraction $x$ ($1-x$), the modified wave function is written as
\begin{eqnarray}
\bar\phi_M({\bf b}',{\bf b})&\equiv&
\int \frac{d^2 {\bf k}'_T}{(2\pi)^2}\frac{d^2 {\bf k}_T}{(2\pi)^2}\exp(-i{\bf k}'_{T}
\cdot {\bf b}')\exp(-i{\bf k}_{T}\cdot {\bf b})\bar\phi_M({\bf k}'_{T},{\bf k}_{T}),\nonumber\\
&=&\frac{2\beta_M^2}{\pi}\phi_M(x) \exp\left[-2\beta_M^2xb^{\prime
2}-2\beta_M^2(1-x)b^{2}\right] .\label{6}
\end{eqnarray}

Our goal is to find a function $S({\bf b})$, such that the Glauber effect is large
(small) for $M=\pi$ ($M=\rho$). The similar Glauber factor, describing the medium
effect \cite{LRW07} in Relativistic Heavy Ion Collider physics, respects the normalization $S(0)=0$ \cite{LRW07}.
If $S({\bf b})$ increases with $b$ monotonically, the real piece $\cos[S({\bf b})]$
takes values in both the first and second quadrants for finite $b$, so its
contributions from these two quadrants cancel each other. The contribution from the third
quadrant, i.e., from large $b$, may not be important due to the suppression of the
exponential in Eq.~(\ref{6}). The imaginary piece $\sin[S({\bf b})]$ remains positive
in the first and second quadrants, such that its effect always exists and becomes small
only in the trivial case with $S({\bf b})\to 0$. Therefore, a monotonic function for
$S({\bf b})$, which tends to enhance both the $B^0\to\pi^0\pi^0$ and $\rho^0\rho^0$
branching ratios, is not preferred. A polynomial function
or a sinusoidal function can provide an oscillatory $S({\bf b})$ in $b$. Because
the large $b$ region is suppressed, we can simply parametrize $S({\bf b})$ by
a sinusoidal function
\begin{eqnarray}
S({\bf b})=r\pi\sin(p b), \label{eq:gau}
\end{eqnarray}
where the tunable parameters $r$ and $p$ govern the magnitude and the
frequency of the oscillation, and should take the same values for the pion
and the $\rho$ meson due to the universality of the Glauber factor.

\subsection{Numerical results}

The following $B$ meson wave function~\cite{LUY,KLS} is employed
in the numerical analysis,
\begin{eqnarray}
\phi_B(x,b)&=&N_Bx^2(1-x)^2
\exp\left[-\frac{1}{2}\left(\frac{xm_B}{\omega_B}\right)^2
-\frac{\omega_B^2 b^2}{2}\right],
\end{eqnarray}
with the coefficient $N_B$ being determined through the normalization
condition
\begin{eqnarray}
\int_0^1 dx \phi_{B}(x, b=0) &=& \frac{f_{B}}{2 \sqrt{2N_c}}.
\end{eqnarray}
We take the distribution amplitudes
\begin{eqnarray}
\phi_{\pi}^A(x)&=&\frac{6f_\pi}{2\sqrt{2N_c}}x(1-x)\bigg[1+
\frac{3}{2} a_2^{\pi} \bigg( 5(2x-1)^2  - 1 \bigg)+ \frac{15}{8} a_4^{\pi}
\bigg(1- 14(2x-1)^2 +4 (2x -1)^4\bigg)\bigg],\nonumber\\
\phi_{\pi}^P(x)&=&\frac{f_{\pi}}{2\sqrt{2N_c}}\, \bigg[ 1
+\frac{1}{2} \left(30\eta_3 -\frac{5}{2}\rho_{\pi}^2\right) \bigg(3 (2x-1)^2 -1\bigg) \nonumber\\ &
& \hspace{35mm} -\, \frac{3}{8}\left\{ \eta_3\omega_3 +
\frac{9}{20}\rho_{\pi}^2(1+6a_2^{\pi}) \right\} \bigg(3- 30 (2x-1)^2 +35 (2x-1)^4\bigg)
\bigg]\;,\\
\phi_{\pi}^T&=&\frac{f_{\pi}}{2\sqrt{2N_c}}\,
(1-2x)\bigg[ 1 + 6\left(5\eta_3 -\frac{1}{2}\eta_3\omega_3 -
\frac{7}{20}
      \rho_{\pi}^2 - \frac{3}{5}\rho_{\pi}^2 a_2^{\pi} \right)
(1-10x+10x^2) \bigg],
\end{eqnarray}
for the pion~\cite{Ball99:Pseudoscalar}, and
\begin{eqnarray}
\phi_{\rho}(x)&=&\frac{3f_{\rho}}{\sqrt{6}} x
(1-x)\left[1+ \frac{3}{2}a_{2\rho}^{||} \bigg( 5(2x-1)^2  - 1 \bigg)\right]\;,\label{eq:lda}\\
\phi_{\rho}^T(x)&=&\frac{3f^T_{\rho}}{\sqrt{6}} x
(1-x)\left[1+ \frac{3}{2} a_{2\rho}^{\perp} \bigg( 5(2x-1)^2  - 1 \bigg)\right]\;,\label{eq:tda}\\
\phi^t_\rho(x) &=& \frac{3f^T_\rho}{2\sqrt 6}(2x-1)^2,\;\;\;\;\;\;\;\;\;\;\;
  \hspace*{0.5cm} \phi^s_\rho(x)=-\frac{3f_\rho^T}{2\sqrt 6} (2x-1)~,\\
\phi_\rho^v(x)&=&\frac{3f_\rho}{8\sqrt6}\bigg(1+(2x-1)^2\bigg),\;\;\; \ \ \
 \phi_\rho^a(x)=-\frac{3f_\rho}{4\sqrt6}(2x-1),
\end{eqnarray}
for the $\rho$ meson~\cite{rho,LM06}. The Glauber factor is introduced
only to the dominant longitudinal-polarization
contribution in the $B^0\to\rho^0\rho^0$ decay. This treatment makes sense, since
the Glauber effect is moderate in this mode as shown later.

Before evaluating the $B \to \pi \pi$ and $\rho^0\rho^0$ branching ratios,
we explain the determination of the parameters $\beta_\pi$ and
$\beta_\rho$ in the TMD pion and $\rho$ meson wave functions in Eq.~(\ref{mod3}).
The parameter $\beta_\pi$ around $0.40$ GeV has been widely adopted in
the literature (see for example Ref.~\cite{YXM07}). Due to the suppression from
the additional intrinsic $k_T$ dependence, we lower the shape parameter $\omega_B$
of the $B$ meson wave function from $0.40$ GeV \cite{LMS05} to $0.37$ GeV to maintain
the NLO PQCD result for the $B\to\pi$ transition form factor $F_0^{B \to \pi}$ .
The parameter $\beta_\rho$ is not as well constrained as $\beta_\pi$, and
we find $\beta_\rho \sim \beta_\pi/ 3$ in order to maintain the NLO PQCD result for
the $B\to\rho$ form factor $A_0^{B \to \rho}$. These values of $\beta_\pi$
and $\beta_\rho$ imply that the pion ($\rho$ meson) wave function exhibits a
weaker (stronger) falloff in the parton transverse momentum $k_T$.
This behavior is consistent with the dual role of the pion as a massless NG
boson and as a $q\bar q$ bound state, which requires a tighter spatial
distribution of its leading Fock state relative to higher Fock states \cite{NS08}.
It has been confirmed that the NLO PQCD results for all the $B \to \pi \pi$ and
$\rho \rho$ decay rates are roughly reproduced with the above parameters, the
coefficients $a_2^\pi = 0.115 \pm 0.115$, $a_2^{\rho,||} = 0.10 \pm 0.10$,
$a_2^{\rho,\perp} = 0.20 \pm 0.20$, $a_4^\pi=-0.015$, $\eta_3 =0.015$,
$\omega_3 =-3$, and $\rho_\pi = m_\pi/ m_0^\pi$ with the chiral enhancement factor
$m_0^\pi =1.3$ GeV \cite{LMS05}, and the $\rho$ meson decay constants
$f_\rho=0.216$ GeV and $f_\rho^T=0.165$ GeV \cite{Ball:2006eu}.

\begin{table}[t]
\caption{Branching ratios from the NLO PQCD formalism in units of
$10^{-6}$, in which NLO (NLOG) denotes the results without (with) the Glauber
effect.}\label{br1}
\begin{center}
\begin{tabular}{cccc}
\hline\hline Modes & Data \cite{HFAG,Petric:2014} 
& NLO & NLOG
\\
\hline $B^0 \to \pi^+ \pi^-$ & $ \phantom{0} 5.10 \pm 0.19 $ &
 $\phantom{0}6.19^{+2.09}_{-1.48}(\omega_B)^{+0.38}_{-0.34}(a_2^\pi)$&
 $\phantom{0}5.39^{+1.86}_{-1.31}(\omega_B)^{+0.28}_{-0.25}(a_2^\pi)$
\\
$B^+ \to \pi^+ \pi^0$ & $ \phantom{0}5.48^{+0.35}_{-0.34} $ &
 $\phantom{0}3.35^{+1.08}_{-0.77}(\omega_B)^{+0.23}_{-0.22}(a_2^\pi)$&
 $\phantom{0}4.45^{+1.38}_{-0.99}(\omega_B)^{+0.39}_{-0.36}(a_2^\pi)$
\\
$B^0 \to \pi^0 \pi^0$ & $ \phantom{0}0.90 \pm 0.16 $ &
 $\phantom{0}0.29^{+0.11}_{-0.07}(\omega_B)^{+0.03}_{-0.02}(a_2^\pi)$&
 $\phantom{0}0.61^{+0.16}_{-0.12}(\omega_B)^{+0.14}_{-0.12}(a_2^\pi)$
\\
$B^0 \to \rho^0 \rho^0$ & $ \phantom{0} 0.97 \pm 0.24 $ &
 $\phantom{0}1.06^{+0.29}_{-0.21}(\omega_B)^{+0.19}_{-0.16}(a_2^\rho)$&
 $\phantom{0}0.89^{+0.26}_{-0.18}(\omega_B)^{+0.13}_{-0.10}(a_2^\rho)$
\\
\hline\hline
\end{tabular}
\end{center}
\end{table}

As listed in the column NLO of Table~\ref{br1}, the NLO PQCD results
for the $B^0 \to \pi^0 \pi^0$ and $B^+ \to \pi^+ \pi^0$ branching ratios
without the Glauber effect are much lower than the data, while those of the
$B^0 \to \pi^+ \pi^-$ and $\rho^0 \rho^0$ decays overshoot the central values of
the data. We then implement the Glauber effect, and carefully scan the $r$
and $p$ dependencies of the $B^0\to\pi^0\pi^0$ branching ratio. Two sets of parameters
are selected, $r\sim 0.47$, $p \sim -0.632$ GeV and $r\sim 0.60$, $p \sim 0.544$ GeV,
which give the largest $B^0 \to \pi^0 \pi^0$ branching ratios $0.62 \times 10^{-6}$
and $0.61 \times 10^{-6}$, respectively. For the former, the $B^0 \to \pi^+ \pi^-$,
$B^+ \to \pi^+ \pi^0$ and $B^0 \to \rho^0 \rho^0$ branching ratios are found to be
$5.90 \times 10^{-6}$, $3.88 \times 10^{-6}$, and $1.07 \times 10^{-6}$,
respectively, which deviate from the data. For the latter, we obtain the
$B^0 \to \pi^+ \pi^-$, $B^+\to \pi^+ \pi^0$ and $B^0 \to \rho^0 \rho^0$ branching
ratios $5.39 \times 10^{-6}$, $4.45 \times 10^{-6}$, and $0.89 \times 10^{-6}$,
respectively, presented in the column NLOG of Table~\ref{br1}. These outcomes show the
preferred tendency: the $B^0 \to \pi^+ \pi^-$ and $B^0\to\rho^0 \rho^0$ branching
ratios decrease by 13\% and 16\%, respectively, while the $B^+ \to \pi^+ \pi^0$ and
$B^0 \to \pi^0 \pi^0$ ones increase by 33\% and a factor of 2.1, respectively.
The $B^0 \to\pi^+\pi^-$ branching ratio does not change much, since it is dominated by
the color-allowed tree amplitude $T$, which is less sensitive to the Glauber effect.
The ratio of the enhancement factor
for the $B^0 \to \pi^0 \pi^0$ mode over the reduction factor for the $B^0\to\rho^0 \rho^0$
mode is about 2.5, close to the ratio $3$ derived in Ref.~\cite{Li:2014haa}, where the
Glauber effect is assumed to be negligible in the $B^0 \to \rho^0 \rho^0$ decay.
Varying the shape parameter $\omega_B$ of the $B$ meson wave function
and the Gegenbauer moments $a_2^{\pi, \rho}$ of the pion and $\rho$ meson,
we estimate the theoretical uncertainties in our formalism given in Table~\ref{br1}.
One can see that all our predictions for the branching ratios in the
NLO PQCD formalism with the Glauber effect match the data better.

To have a clear idea of the Glauber effect, we present the amplitudes
${\cal A}(B^0 \to \pi^0\pi^0)$ and ${\cal A}(B^0 \to \rho^0\rho^0)$
(in units of $10^{-2}{\rm GeV}^{3}$) from Figs.~\ref{fig1}(a) and~\ref{fig1}(b),
\begin{eqnarray}
{\cal A}_{a,b}(B^0\to \pi^0\pi^0)&=& \left\{ \begin{array}{lll}
11.86 -i 9.04,& \ \  -7.13+i 6.18,\ \   & {\rm (NLO)}, \\
10.80 -i 7.25,& \ \   7.67 - i 3.42,\ \ &  {\rm (NLOG)},\\
\end{array} \right.\label{pipi}\\
{\cal A}_{a,b}(B^0\to \rho^0\rho^0)&=& \left\{ \begin{array}{lll}
-42.44 + i 24.42, & \ \  28.88 - i 18.07,\ \ & {\rm (NLO)}, \\
-5.78 + i 4.32,& \ \   -3.61 - i 3.23,\ \ &  {\rm (NLOG)},\\
\end{array} \right.\label{amprr}
\end{eqnarray}
respectively, associated with the four-fermion operator $O_2$ (they are not the full
spectator amplitudes). Equation~(\ref{pipi})
indicates that the result of Fig.~\ref{fig1}(a) varies
a bit because of the approximate cancellation of the Glauber phases
associated with the $M_1$ and $M_2$ mesons, as shown in Eq.~(\ref{cmfp}).
The result of Fig.~\ref{fig1}(b) is modified by the Glauber effect significantly
with a sign flip, in agreement with what was found in Ref.~\cite{Li:2014haa}.
It is obvious that the destructive interference between Figs.~\ref{fig1}(a) and
\ref{fig1}(b) has been turned into a constructive one for the $B^0\to \pi^0\pi^0$ decay.
The consequence is that their sum changes from $5.53e^{-i0.54}\times 10^{-2}$ GeV$^3$
in the NLO PQCD approach to $21.33e^{-i0.52}\times 10^{-2}$ GeV$^3$ in the NLO PQCD
approach with the Glauber effect. As for the $B^0\to \rho^0\rho^0$ decay, the broad
distribution of the $\rho$ meson wave function in $b$ space allows cancellation
to occur, which is attributed to the oscillation of the Glauber phase factor.
This is the reason why each amplitude from Figs.~\ref{fig1}(a) and \ref{fig1}(b)
reduces as shown in Eq.~(\ref{amprr}). However, the sum of the two amplitudes
does not change much relative to the change in the $B^0 \to \pi^0\pi^0$ case.
We have examined the sensitivity of the $B^0\to\rho^0\rho^0$ branching ratio
to $r$ and $p$, and confirmed that the predicted branching ratio is quite stable as long
as $p>0.5$ GeV, varying within only 5\%. It is likely that the
leading Fock state of the pion is tight enough to reveal the Glauber effect from
the oscillatory phase factor, while other hadrons with broad spatial distributions cannot.
We might have found plausible explanations for the dynamical origin
of the Glauber phase and for the unique role of the pion mentioned before.

\section{CONCLUSION}

In this paper we have performed the model estimate of the
Glauber effects in the $B\to\pi\pi$ and $\rho\rho$ decays in the PQCD
approach based on the $k_T$ factorization theorem.
The Glauber phase factor, arising from the factorization and all-order
summation of the Glauber gluons for two-body hadronic $B$ meson decays,
is universal as shown in our previous work.
Despite being universal, the Glauber factor does make distinct
impacts on the $B^0\to\pi^0\pi^0$ and $B^0\to\rho^0\rho^0$ branching ratios
through its convolution with the TMD pion and $\rho$ meson TMD wave functions
with different intrinsic $k_T$ dependencies. It was noticed that the pion
($\rho$ meson) wave function exhibiting a weak (strong) falloff in $k_T$
serves the purpose. These behaviors are consistent with the dual role of the
pion as a massless NG boson and as a $q\bar q$ bound state, which requires a
tighter spatial distribution of its leading Fock state relative to higher
Fock states. It has been pointed out that the tight leading Fock state of
the pion may be able to reveal the Glauber effect from the oscillatory phase
factor as parametrized in Eq.~(\ref{eq:gau}), while other hadrons with
broad spatial distributions cannot.

We have demonstrated that the $B^0\to\pi^0\pi^0$ branching ratio is
enhanced by a factor of 2.1, reaching $0.61\times 10^{-6}$, while
the $B^0\to\rho^0\rho^0$ one remains around $0.89\times 10^{-6}$, down
by only 16\%. This observation supports the fact that the Glauber effect from the pion
can be more significant, as postulated in \cite{LM11,Li:2014haa}.
The $B^0\to\pi^+\pi^-$ ($B^+\to\pi^+\pi^0$) branching ratio is modified into
$5.39\times 10^{-6}$, decreasing by 13\% ($4.45\times 10^{-6}$, increasing by 33\%),
such that the consistency between the NLO PQCD predictions and the data is improved
for all the modes. The above changes are due to the facts that the Glauber
phase enhances the color-suppressed tree amplitude by turning the destructive
interference between the LO spectator diagrams into a constructive one, and that it
also modifies the interference between the color-suppressed and color-allowed
tree amplitudes. We stress that the $B\to\pi\pi$ puzzle must be resolved by resorting
to a mechanism that can differentiate the pion from other mesons, and
that the Glauber gluons should be one of the most crucial mechanisms.

\begin{acknowledgments}

X.L. thanks Institute of Physics, Academia Sinica for the warm hospitality during his
visit, where this work was finalized. We thank S.J.~Brodsky, H.Y.~Cheng, C.K.~Chua,
T.~Huang, S.~Mishima, and X.G.~Wu for useful discussions. This work was supported in part
by the Ministry of Science and Technology of R.O.C. under Grant No.
NSC-101-2112-M-001-006-MY3, by National Science Foundation
of China under Grants No.~11205072, No.~10975074, and No.~11235005, and by the Priority Academic
Program Development of Jiangsu Higher Education Institutions (PAPD).

\end{acknowledgments}

\end{document}